# Quadratic Sinusoidal Analysis of Neurons in Voltage Clamp

*Christophe Magnani and L.E.Moore*

CESeM - CNRS UMR 8194 - Université Paris Descartes
45 rue des Saints-Pères - 75270 PARIS Cedex 6

**Abstract**

Nonlinear biophysical properties of individual neurons are known to play a major role in the nervous system. Earlier electrophysiological studies have made use of piecewise linear characterization of voltage clamped neurons, which consists of a sequence of linear admittances computed at different voltage levels. In this paper, the linear approach is extended to a piecewise quadratic characterization in two different ways. First, an analytical model is derived with power series following the work pioneered by Fitzhugh. Second, matrix calculus is developed to provide a novel quantitative analysis not dependent on differential equations. This method provides an assessment of quadratic responses for both data recorded from individual neurons and their corresponding models.

## 1 INTRODUCTION

In an innovative paper, FitzHugh (1983) derived analytically the nonlinear response to a single sinusoidal stimulation for the voltage clamp Hodgkin and Huxley (1952) model. He showed that the steady-state current response to a single sinusoidal frequency $f$ has harmonic components $f, 2f, 3f, \ldots$ This analysis provided a quantitative interpretation of the harmonic components observed by Moore et al. (1980).

However, the single sinusoidal stimulation is generally insufficient to characterize the nonlinear behavior in neurons. In particular, it is unable to predict the quadratic response to a double sinudoidal stimulation of frequencies $f_1, f_2$ since additional intermodulation products $f_1 + f_2$ and $|f_1 - f_2|$ occur in the measured membrane current. In this paper, a quadratic approximation of the neuronal response to a double sinudoidal stimulation is derived analytically and extended to a multi-sinusoidal stimulation.

This approach leads to the development of a *Quadratic Sinusoidal Analysis*, or more concisely QSA, relying on matrix calculus and eigendecomposition in order to characterize the nonlinear behavior from neuronal responses to a multi-sinusoidal stimulation in the steady state. This novel method is devoted to exploring in depth the behavior of individual neurons, which is fundamentally nonlinear and cannot be described by linear theory alone.

Finally, the QSA is applied to electrophysiological experiments by using Volterra series techniques, providing an extension of the earlier piecewise linear analysis (Fishman et al., 1977 ; Murphey et al., 1995) to a piecewise quadratic analysis.

## 2 THEORY

### 2.1 Double sinusoidal voltage clamp

The proposed model implements a minimal soma with only one kinetic equation in order to simplify the calculations while preserving their physiological significance. The parameters were selected to be consistent with experimental data published in Idoux et al. (2008).

$$CV' = I - I_L - I_K - I_{Na} \qquad (2.1)$$
$$n' = \alpha_n (1-n) - \beta_n n \qquad (2.2)$$

Here $I_L = \overline{g_L}(V - V_L)$, $I_K = \overline{g_K} n (V - V_k)$ and $I_{Na} = \overline{g_{Na}} m_\infty (1-n)(V - V_{Na})$ represent the leakage, $K^+$ and $Na^+$ ionic currents respectively. $V$ is the imposed membrane potential, $I$ the measured current, $n$ the gating variable for $K^+$, $m_\infty$ the gating variable for $Na^+$ at equilibrium ($m' = 0$). The other values are constant parameters : the membrane capacitance $C = 0.0000205\mu$F, the maximal conductances $\overline{g_L} = 0.00137\mu$S, $\overline{g_K} = 0.00118\mu$S, $\overline{g_{Na}} = 0.00064\mu$S and the reversal potentials $V_L = -53$mV, $V_K = -87$mV, $V_{Na} = 77$mV for leakage, $K^+$ and $Na^+$ respectively. The functions $\alpha_n$ and $\beta_n$ depend on the variable $V$ and their mathematical expressions are fully described by Murphey et al. (1995) for $vm = -35$mV, $sm = 0.056$mV$^{-1}$, $vn = -39$mV, $sn = 0.09$mV$^{-1}$, $tn = 0.1$s when defining $\alpha_n = e^{2s_n(V-v_n)}/(2t_n)$, $\beta_n = e^{-2s_n(V-v_n)}/(2t_n)$ and $m_\infty = 1/\left(1 + e^{-4s_m(V-v_m)}\right)$.

Fitzhugh imposes a single sinusoidal command for the membrane potential $V(t) = V_0 + V_1 \cos(\omega_1 t)$ where $V_0$ is the DC constant, $V_1$ is the amplitude and $\omega_1$ is the angular frequency. This can be extended to a double sinusoidal command

$$V(t) = V_0 + V_1 \cos(\omega_1 t + \phi_1) + V_2 \cos(\omega_2 t + \phi_2) \qquad (2.3)$$

where $V_1, V_2$ are the amplitudes, $\omega_1, \omega_2$ are the angular frequencies and $\phi_1, \phi_2$ the phases. The phase difference $|\phi_2 - \phi_1|$ is especially important to ensure that the two sine waves are uncorrelated. Also it is necessary to have $\omega_1$ and $\omega_2$ distinct to avoid degenerate cases. The notations can be simplified by putting $c_1 = \cos(\omega_1 t + \phi_1)$, $c_2 = \cos(\omega_2 t + \phi_2)$, $s_1 = \sin(\omega_1 t + \phi_1)$ and $s_2 = \sin(\omega_2 t + \phi_2)$.

The goal is to determine an analytical expression for the current $I$. In this paper, the solutions are limited to a quadratic approximation, which is the minimal degree of nonlinearity. Indeed, although neuronal responses generally show higher degrees of nonlinearity, they can be ignored if the stimulation amplitude is sufficiently small.

The gating variable $n$ can be approximated by a quadratic polynomial near the steady state $n_0$ with respect to the input fluctuations amplitudes $V_1$ and $V_2$ :

$$n \simeq n_0 + V_1 n_1 + V_2 n_2 + V_1^2 n_{11} + V_2^2 n_{22} + V_1 V_2 n_{12} \qquad (2.4)$$

where $n_1, n_2, n_{11}, n_{22}, n_{12}$ are unknown functions of time. Similarly, $\alpha_n$ and $\beta_n$ can be approximated by quadratic polynomials with respect to $V_1$ and $V_2$ after a quadratic Taylor decomposition :

$$\begin{aligned}
\alpha_n &\simeq \alpha_n(V_0) \\
&+ \alpha_n'(V_0)(V_1 c_1 + V_2 c_2) \\
&+ \frac{\alpha_n''(V_0)}{2}(V_1 c_1 + V_2 c_2)^2
\end{aligned}$$

$$\begin{aligned}
\beta_n &\simeq \beta_n(V_0) \\
&+ \beta_n'(V_0)(V_1 c_1 + V_2 c_2) \\
&+ \frac{\beta_n''(V_0)}{2}(V_1 c_1 + V_2 c_2)^2
\end{aligned}$$

The approximated expressions of $n, \alpha_n, \beta_n$ are polynomials in variables $V_1, V_2$, which can be substituted into the equation 2.2. Then, by identification with the zero polynomial, the system reduces to a set of five linear differential equations as well as the common steady-state expression $n_0 = \frac{\alpha_0}{\alpha_0 + \beta_0}$ :

$$\begin{aligned}
n_1' + \lambda n_1 + A c_1 &= 0 \\
n_2' + \lambda n_2 + A c_2 &= 0 \\
n_{11}' + \lambda n_{11} + B_1 c_1^2 + C_1 c_1 s_1 &= 0 \\
n_{22}' + \lambda n_{22} + B_2 c_2^2 + C_2 c_2 s_2 &= 0 \\
n_{12}' + \lambda n_{12} + D_{12} c_1 c_2 + E_1 c_2 s_1 + E_2 c_1 s_2 &= 0
\end{aligned}$$

where $\lambda = \alpha_n(V_0) + \beta_n(V_0)$ is constant, $A$ is constant, and $B_1(\omega_1)$, $B_2(\omega_2)$, $C_1(\omega_1)$, $C_2(\omega_2)$, $D_{12}(\omega_1, \omega_2)$, $E_1(\omega_1)$, $E_2(\omega_2)$ are rational functions. The details of these cumbersome expressions are not important, except for their frequency content. From trigonometric calculus, $c_1, c_2, c_1^2, c_2^2$ contain frequencies $\omega_1$, $\omega_2$, $2\omega_1$, $2\omega_2$ respectively, and $c_1 c_2$, $c_2 s_1$, $c_1 s_2$ contain $|\omega_1 \pm \omega_2|$. The five differential equations being linear, their stationary solutions must preserve the frequencies, namely the functions $n_1$, $n_2$, $n_{11}$, $n_{22}$, $n_{12}$ are associated with the frequencies $\omega_1$, $\omega_2$, $2\omega_1$, $2\omega_2$, $|\omega_1 \pm \omega_2|$ respectively. This remark is the fundamental principle of the QSA method, namely these are the response frequencies that characterize the nonlinear behavior. These differential equations were solved by MATHEMATICA 7 (Wolfram Research, Champaign, IL, USA) after transformation into algebraic equations by the Laplace transform. The transient terms like $e^{-t(\alpha_n(V_0)+\beta_n(V_0))}$ were ignored to retain only stationary solutions.

The current $I$ can also be approximated by a quadratic polynomial near the steady state $I_0$ with respect to the input amplitudes $V_1$ and $V_2$

$$I \simeq I_0 + V_1 I_1 + V_2 I_2 + V_1^2 I_{11} + V_2^2 I_{22} + V_1 V_2 I_{12} \tag{2.5}$$

The expressions of $I_1, I_2, I_{11}, I_{22}, I_{12}$ are directly determined by polynomial identification from equation 2.1 after substitution of $n$ by its quadratic polynomial approximation 2.4. Similarly, $I_1$, $I_2$, $I_{11}$, $I_{22}$, $I_{12}$ are associated with the frequencies $\omega_1$, $\omega_2$, $2\omega_1$, $2\omega_2$, $|\omega_1 \pm \omega_2|$ respectively.

## 2.2 Multi-sinusoidal voltage clamp

For a single sinusoidal voltage clamp, the frequency space is described by one variable $\omega_1$. For a double sinusoidal voltage clamp, the frequency space is described by two variables $\omega_1$ and $\omega_2$. If each variable describes $N$ frequencies, then $\frac{1}{2}N(N+1)$ pairs of frequencies are required to probe the quadratic neuronal response. For instance, $N = 10$ would require 55 experiments for only one voltage level and stimulus amplitude. This would be experimentally not reasonable due to an excessively long recording duration for a whole cell voltage clamped neuron.

A solution consists of computing the quadratic response for all pairs in parallel instead of sequentially. For this, the double sinusoidal command has to be extended to a multi-sinusoidal command as follows

$$V(t) = V_0 + \sum_{i=1}^{N} V_i \cos(\omega_i t + \phi_i) \tag{2.6}$$

The quadratic polynomials in two variables 2.4 and 2.5 have to be extended to quadratic polynomials in several variables. The current response is then approximated by

$$I \simeq I_0 + \sum_{i=1}^{N} V_i I_i + \sum_{i=1}^{N} V_i^2 I_{ii} + \sum_{i=1}^{N} \sum_{j=i+1}^{N} V_i V_j I_{ij} \tag{2.7}$$

The coefficients $I_i, I_{ii}, I_{ij}$ are determined by polynomial identification as in the previous section. In particular, it can be checked that for all $V_i = 0$ except $V_k \neq 0$ and $V_l \neq 0$ ($k \neq l$) the multi-sinusoidal current 2.7 coincides with the double sinusoidal current 2.5. In practice however, the analytical formula 2.7 is more simply constructed by looping the double sinusoidal voltage clamp over all the frequency pairs $\{i, j\}$. A result of this algorithm is illustrated (figure 2.1) for the frequency set $\{0.2, 0.8, 2, 3.4, 5.8, 10.4, 13.4, 17.8\}$ (in Hertz) with sinusoidal amplitudes equal to 0.5 mV and randomized phases. The voltage command has a mean of $V_0 = -43$ mV and standard deviation 0.99 mV. The quadratic terms of $I(t)$ are required to accurately describe the neuronal response, which clearly cannot be done by the usual linear analysis.

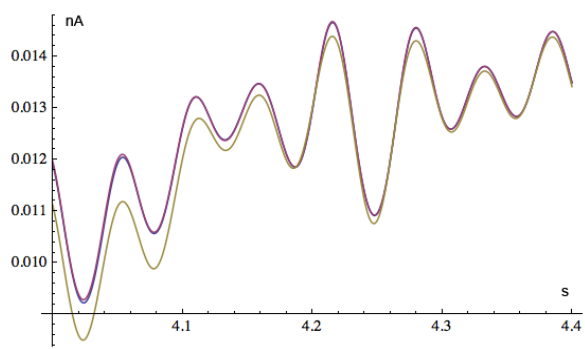

Fig. 2.1: Superposition of the original current response (in blue), the quadratic analysis (in red) and the linear analysis (in green). The red curve is almost perfectly superimposed to the blue curve. Clearly, the quadratic analysis is required to accurately describe the neuronal response.

## 2.3 Linear and quadratic behavior

The multi-sinusoidal voltage clamp formulas 2.6 and 2.7 can be rewritten in matrix form in order to simplify the calculations, and further to analyse experiments. It is well known from Fourier analysis that complex exponentials are optimal to represent stationary signals. More precisely, for an experiment of duration $T$ (in seconds), the elementary wave functions $\mathbf{e_k}(t) = e^{i2\pi kt/T}$ are able to reconstruct $V(t)$ and $I(t)$ by linear superposition. The stimulation frequencies being integer multiples of $2\pi/T$, they can be denoted by $\omega_i = 2\pi n_i/T$ where $i$ is an index describing the set $\Gamma = \{-N, \ldots, -1, +1, \ldots, +N\}$. Also, by convention $n_{-i} = -n_i$. The multi-sinusoidal voltage command can be directly written as a superposition of elementary waves through the common trigonometric formula $cos(\theta) = \frac{1}{2}\left(e^{i\theta} + e^{-i\theta}\right)$

$$\mathbf{v} = V - V_0 = \sum_{k \in \Gamma} v_k \mathbf{e_{n_k}} \qquad (2.8)$$

where $v_k = \frac{1}{2}V_k e^{i\phi_k}$ for $k > 0$ and $v_{-k} = \overline{v_k}$ (bar is complex conjugate). In fact, this expression represents the multi-sinusoidal voltage command as a vector $\mathbf{v}$ with components $v_k$ in the basis of elementary waves $\mathbf{e_{n_k}}$.

The linear part of the current response $\mathbf{i_1} = \sum_{i=1}^{N} V_i I_i$ involves stimulation frequencies only and thus can be written as a linear superposition of the elementary waves with complex coefficients $L_k$ acting on the input like an admittance

$$\mathbf{i_1} = \sum_{k \in \Gamma} L_k v_k \mathbf{e_{n_k}} \tag{2.9}$$

By contrast, the quadratic part of the current response $\mathbf{i_2} = \sum_{i=1}^{N} V_i^2 I_{ii} + \sum_{i=1}^{N} \sum_{j=i+1}^{N} V_i V_j I_{ij}$ involves frequencies $2\omega_i$ and $|\omega_i \pm \omega_j|$. Therefore, products $\mathbf{e_{n_i}} \mathbf{e_{n_j}}$ are produced such that the quadratic response can be written as a quadratic mixing of the elementary waves

$$\mathbf{i_2} = \sum_{i \in \Gamma} \sum_{j \in \Gamma} B_{i,j} v_i v_j \mathbf{e_{n_i}} \mathbf{e_{n_j}} \tag{2.10}$$

In order to ignore constant DC in the pure quadratic response, the coefficients $B_{i,-i}$ must be set to zero. Moreover, since the current response has no imaginary part, the coefficients must satisfy $B_{i,j} = \overline{B_{-i,-j}}$. Also, note the symmetry $B_{i,j} = B_{j,i}$.

Remarkably, the row flipped matrix $Q_{i,j} = B_{-i,j}$ is Hermitian (Lang, 2002). This is very convenient because Hermitian matrices have many important properties. In particular, their eigenvectors can be used to decompose the quadratic current response as a sum of squares weighted by real eigenvalues playing the role of amplitudes. The general skeleton of $Q_{i,j}$ is as follows

$$\begin{pmatrix} 0 & \ldots & B_{N,-1} & B_{N,1} & \ldots & B_{N,N} \\ \vdots & \ldots & \ldots & \ldots & \ldots & \vdots \\ B_{1,-N} & \ldots & 0 & B_{1,1} & \ldots & B_{1,N} \\ B_{-1,-N} & \ldots & B_{-1,-1} & 0 & \ldots & B_{-1,N} \\ \vdots & \ldots & \ldots & \ldots & \ldots & \vdots \\ B_{-N,-N} & \ldots & B_{-N,-1} & B_{-N,1} & \ldots & 0 \end{pmatrix}$$

This matrix is the essential tool of the method and is called the *QSA Matrix*. It is very appropriate for computations with the Fourier transform, as explained in the following section on experimental measurements.

This matrix allows the reconstruction of the current response through simple algebraic manipulations. Indeed, if the clock matrix is defined by $U_t = \text{diag}(\mathbf{e_{n_k}}(t))$ then the voltage command vector as well as the linear and quadratic transformations $L$ and $Q$ can be made explicitly dependent of the time, namely $\mathbf{v_t} = U_t \mathbf{v}$ and $L_t = L U_t$ and $H_t = U_t^* Q U_t$ (the upper $^*$ denotes the conjugate transpose). This allows a reconstruction of the current response in the time domain by considering $L$ as a linear form and $Q$ as a Hermitian form

$$I(t) - I_0 \simeq L\mathbf{v_t} + \mathbf{v_t}^* Q \mathbf{v_t}$$

or equivalently

$$I(t) - I_0 \simeq L_t \mathbf{v} + \mathbf{v}^* Q_t \mathbf{v}$$

It is interesting to note the duality of these two formulations, analogous to the Schrödinger / Heisenberg pictures in quantum mechanics. Indeed, either the vector is time-dependent and operators are time-independent, or the converse.

The QSA matrix being Hermitian, it can be diagonalized through $Q = P^* D P$ where $P$ is a unitary matrix satisfying $P^* = P^{-1}$. In this expression, each column in $P^*$ contains the coordinates of an eigenvector expressed in the basis of elementary waves. Also, $D = \text{diag}(d_i)$ is the diagonal matrix containing the eigenvalues. The quadratic part can then be rewritten as

$$\mathbf{i_2}(t) = \mathbf{v_t}^* Q \mathbf{v_t} = \mathbf{w_t}^* D \mathbf{w_t}$$

where $\mathbf{w_t} = P \mathbf{v_t}$. The transformation matrix $P$ being unitary, it preserves the signal energy of the stimulation vector, namely $\|\mathbf{w_t}\|^2 = \|\mathbf{v_t}\|^2$. On the other hand, the diagonal matrix $D$ plays the role of a quadratic filter such that in the above change of basis the quadratic part of the response is reduced to a sum of squares

$$\mathbf{i_2}(t) = \mathbf{v_t}^* Q \mathbf{v_t} = \sum_{i \in \Gamma} d_i |\mathbf{w_t}|_i^2$$

This reduction has a special meaning when only few eigenvalues are dominant. In this case, the neuronal function can be approximated by ignoring the small eigenvalues, providing a more compact description. However, the total contribution of all the eigenvalues is equal to zero because $\sum d_i = \text{Tr}(D) = \text{Tr}(Q) = 0$. The QSA matrix and the eigenvalues of the model are illustrated (2.2), showing two dominant eigenvalues. The computations were made with MATLAB (The MathWorks, Natick, MA, USA).

## 3 RESULTS AND DISCUSSION

### 3.1 Nonoverlapping measurements

In practice, experimental measurements are subject to difficulties due to frequency overlapping. More precisely, it is possible that $n_i + n_j = n_k + n_l$ for distinct pairs of frequencies $\{n_i, n_j\}$ and $\{n_k, n_l\}$. In this case, the terms $B_{i,j} v_i v_j \mathbf{e_{n_i}} \mathbf{e_{n_j}}$ and $B_{k,l} v_k v_l \mathbf{e_{n_k}} \mathbf{e_{n_l}}$ share the same output component $\mathbf{e_{n_i+n_j}} = \mathbf{e_{n_k+n_l}}$. This means that the measurement of such a shared component is unable to separate the coefficients $B_{i,j}$ and $B_{k,l}$. This problem is general and also encountered when measuring Volterra kernels in nonlinear signal theory.

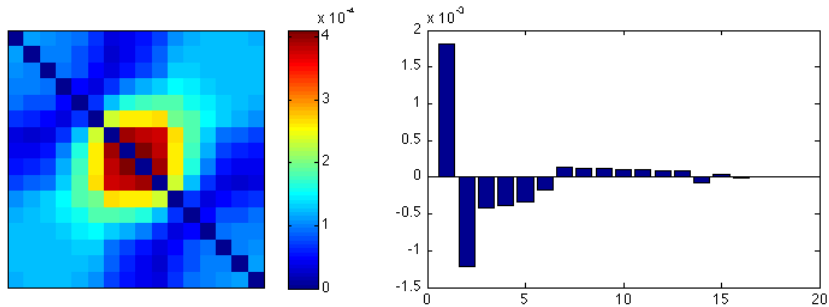

Fig. 2.2: Magnitude of the QSA matrix (left) and its eigenvalues (right) for the model. The eigenvalues have been sorted by decreasing magnitude. There are two dominant eigenvalues suggesting that the quadratic neuronal function can be considered as a sum of two squares at first approximation. For these plots, the frequency components were computed with the MATLAB command FFT divided by the number of points.

Harmonic probing has been developed as a practical measurement technique to determine the kernels in the frequency domain (see especially Boyd et al., 1983). For instance, when a multi-sinusoidal voltage command is imposed with incommensurable frequencies $\omega_1, \ldots, \omega_N$ then every coefficient of the corresponding second order Volterra kernel $G_2(\omega_i, \omega_j)$ can be deduced from the output measured at $\omega_i + \omega_j$. Other methods have also made use of incommensurable frequencies, such as Victor and Shapley (1980). In this paper, harmonic probing was adapted to determine the coefficients of the QSA matrix without frequency overlapping. In particular, a flexible algorithm was developed to generate sets of nonoverlapping frequencies appropriate for the voltage clamp conditions (controlled duration and frequency range).

Then, for a set of nonoverlapping frequencies, the equation 2.10 can be solved in which $B_{i,j}$ are the unknown coefficients.

$$B_{i,j} = \gamma_{i,j} \frac{\hat{I}(n_i + n_j)}{v_i v_j}$$

where $\hat{I}(n_i + n_j)$ coincides with the Fourier component of $I(t)$ at the frequency $\omega_i = 2\pi(n_i + n_j)/T$. The term $\gamma_{i,j} = \frac{1}{2} + \frac{1}{2}\delta_{i,j}$ is a coefficient of symmetry such that $\gamma_{i,i} = 1$ and $\gamma_{i,j} = \frac{1}{2}$ for $i \neq j$, which implies $B_{i,i} = \frac{\hat{I}(2n_i)}{v_i^2}$ and $B_{i,j} = B_{j,i} = \frac{1}{2}\frac{\hat{I}(n_i + n_j)}{v_i v_j}$ respectively.

## 3.2 Analysis of prepositus hypoglossi neurons

An extension of linear piecewise analysis (Fishman et al., 1977 and Murphey et al., 1995) to quadratic piecewise analysis was developed by using the QSA matrix. As explained above, a nonoverlapping QSA is required in order to experimentally obtain a nonlinear characterization of neuronal behavior. For this purpose, voltage clamp data of a prepositus hypoglossi neuron provided by Professor Daniel Eugène (personal communication) were analyzed using nonoverlapping frequencies $\{0.2, 0.8, 2, 3.4, 5.8, 10.4, 13.4, 17.8\}$ (in Hertz) at two voltage levels $-60$ mV and $-55$ mV. A rectangle low-pass filter was applied a posteriori to remove noise greater than 36 Hz. The highest stimulation frequency is 17.8 Hz which implies that the highest frequency of the quadratic response is $2 * 17.8 = 35.6$ Hz, hence the cutoff at 36 Hz is valid for a quadratic analysis. Figure 3.1 represents the current responses in the time domain. Clearly, the quadratic response is more accurate than the linear one. The residual error is due to experimental noise or higher order frequency contamination, which is inevitable in any experiment. Each data sequence for the two experiments is an average of 4 recordings or more using the experimental protocol as described in Idoux et al. (2008). The adequacy of the quadratic analysis was observed in all experiments, except when the stimulation amplitude is either too large evoking higher order frequency contamination or too small to overcome synaptic or intrinsic channel fluctuations.

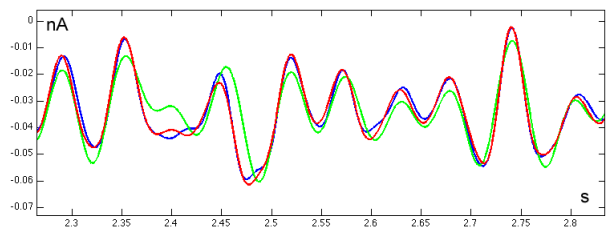

Fig. 3.1: Experimental current response $I(t)$ to a stimulation $V(t)$ with nonoverlapping frequencies centered at $-55.26 \pm 2.85$ mV. The quadratic analysis (in red) is dramatically more accurate than the linear analysis (in green) to describe the experimental response (in blue). Indeed, the red curve is almost superimposed on the blue curve.

One of the most important condition to ensure the quality of a nonlinear voltage clamp experiment is the time invariance, which means that the same voltage input must always generate the same current output. Therefore, it is necessary to compute the correlation between all the recordings in order to ensure that they are reasonably time invariant. The quadratic response $\mathbf{i_2}(t)$ (defined previously) can be extracted from the full response $I(t)$ by Fourier analysis for each of the $M$ recordings, in this experiment $M = 4$. This provides $M$ signals $r_1 = \mathbf{i_{2,1}}(t), \ldots, r_M = \mathbf{i_{2,M}}(t)$. In fact, each $\mathbf{i_{2,m}}$ is the quadratic part of the $m$-th recording. The pairwise correlations are then encoded into the matrix of pairwise products $\langle r_i, r_j \rangle = \int r_i(t) r_j(t) dt$

$$\begin{pmatrix} \langle r_1, r_1 \rangle & \langle r_1, r_2 \rangle & \ldots & \langle r_1, r_M \rangle \\ \langle r_2, r_1 \rangle & \langle r_2, r_2 \rangle & \ldots & \langle r_2, r_M \rangle \\ \vdots & \ldots & \ldots & \vdots \\ \langle r_M, r_1 \rangle & \langle r_M, r_2 \rangle & \ldots & \langle r_M, r_M \rangle \end{pmatrix}$$

The symmetry of the matrix reduces the number of computations. From this, the time invariance correlation coefficient *ticc* can be defined as the coefficient of variation of the elements of this matrix, that is $ticc = \sigma/\mu$ where $\sigma$ and $\mu$ are the mean and the standard deviation of the elements of this matrix. When all dot products are identical the *ticc* is zero, otherwise it increases depending on the lack of correlation. Although empirical, the *ticc* has proved to be particulary efficient to make automatic data extraction from large pools of experiments. In particular, the criterion $ticc < 1$ was used for the analyzed experiments. For the two experiments, the *ticc* is 0.5989 at $-60$mV and 0.1087 at $-55$mV.

The figures 3.2 A and B compare the magnitudes of the QSA matrices. In general, the magnitudes tend to globally increase at depolarized levels, as illustrated here when comparing $-60$mV to $-55$mV. The figures 3.2 C and D compare the magnitudes of the interpolated QSA matrices. The interpolations were performed by the MATLAB command GRIDDATA (linear method) in order to represent the responses in 3D color plots over a continuous range of frequencies. The approach allows a coarse approximation of the response including overlapping frequencies. This can be further improved by combining additional QSA matrices constructed from other nonoverlapping measurements. As can be observed, the peaks of the frequency interactions are approximately in the same location after depolarization. Moreover, new peaks appear at high frequencies.

The figure 3.3 compares the linear and quadratic analyses for the two experiments. An interesting result is that the first eigenvalue is dominant, especially at $-55$mV. This means that a single large square $d_i \left| \mathbf{w_t} \right|_i^2$ plays a major role in the description of the neuronal function. However, at different membrane potentials or other types of neurons there may be two or more significant eigenvalues. It would appear that the quadratic neuronal function provides an indication of the complexity of the information processing used by neurons. An important observation is that the eigenvalues increase at the depolarized levels consistent with the increased QSA matrix amplitudes.

The last plot of the figure 3.3 shows a summation by columns of the QSA matrix

$$R(j) = \sum_{i \in \Gamma} |Q_{i,j}|$$

Hence, each value $R(j)$ represents the quadratic interactions involving the stimulation frequency $\omega_j$. The advantage of this summation is that a Bode-like plot can be made. Again, the magnitude is larger at the depolarized level, which is consistent with and confirms the eigenvalue analysis. This plot shows that each stimulation frequency can significantly contribute to the nonlinear response. Clearly, $R(j)$ has been enhanced during the voltage clamped depolarization at the higher frequencies as shown in the figure 3.3.

In conclusion, the QSA matrix is a novel method to characterize the nonlinear responses of individual neurons. The eigendecomposition allows an intrinsic description of the neuron's quadratic function. Furthermore, the complexity inherent to the nonlinear neuronal behavior as illustrated by the analytical expressions derived in this paper, is dramatically simplified by the QSA approach. In particular, the dominance of a few eigenvalues can provide a very compact representation of an individual complex neuron. This suggests new possibilities for large scale neural network simulations.

# 4 ACKNOWLEDGMENTS

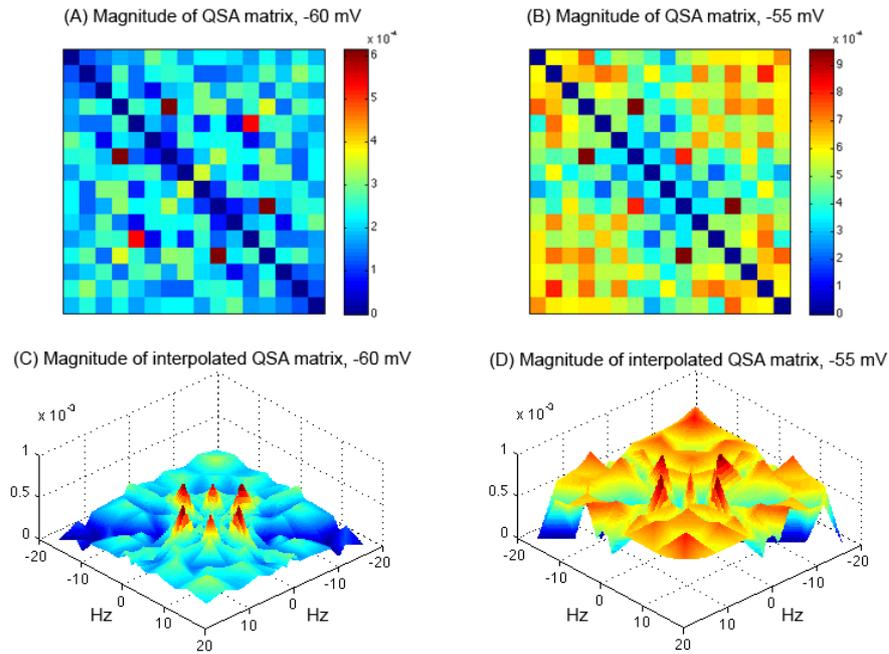

Fig. 3.2: Analysis of the experimental data of a prepositus hypoglossi neuron at
−60mV (left column) and at −55mV (right column)

(A and B) Magnitude of the QSA matrix computed from the two experiments at −60mV (A) and at −55mV (B). Clearly, the coefficients (color coded) are increased at the depolarized level.

(C and D) Magnitude of the interpolated QSA matrix at −60mV (C) and −55mV (D). The peaks of the frequency interactions are approximately in the same location after depolarization, moreover additional peaks appear at high frequencies.

For these plots, the frequency components were computed with the MATLAB command FFT divided by the number of points. The interpolations were performed by the MATLAB command GRIDDATA (linear method). The current was measured in nA and the voltage in mV.

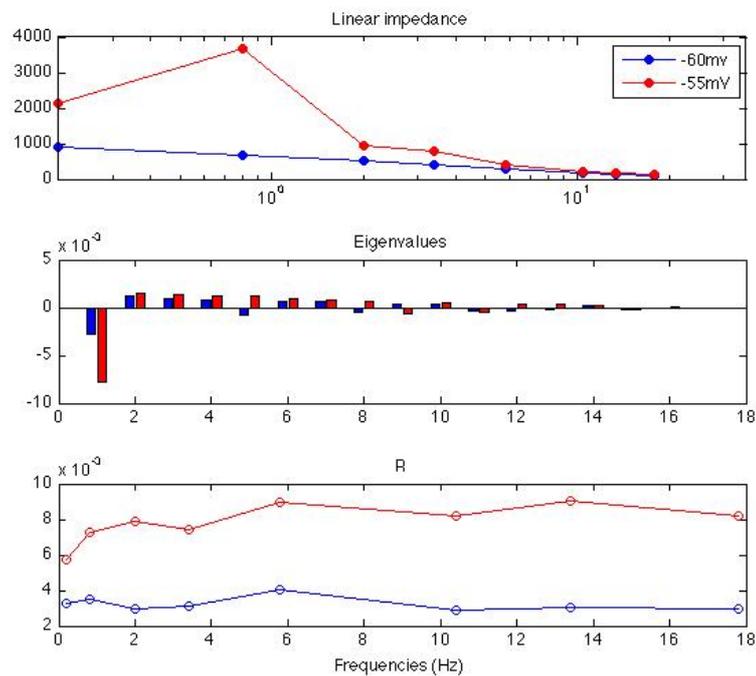

Fig. 3.3: Linear and nonlinear analysis of the two experiments. At the top, the impedance computed from usual linear analysis. At the middle, the eigenvalues of each QSA matrix. At the bottom, the R summation of each QSA matrix. For these plots, the frequency components were computed with the MATLAB command FFT divided by the number of points. The current was measured in nA and the voltage in mV.